# ASSEMBLY ASM291031v2 (GENBANK: GCA_002910315.2) IDENTIFIED AS ASSEMBLY OF THE NORTHERN DOLLY VARDEN (*Salvelinus malma malma*) GENOME, AND NOT THE ARCTIC CHAR (*S. alpinus*) GENOME

S.V. Shedko

*Federal Scientific Center of the East Asia Terrestrial Biodiversity,*

*Far Eastern Branch of the Russian Academy of Sciences, Vladivostok, 690022*

*Russia*

*e-mail: shedko@biosoil.ru*

To date, twelve complete genomes representing eleven species belonging to six genera have been sequenced in salmonids. For the genus *Salvelinus*, it was supposed to sequence the genome of Arctic char, one of the most variable species of vertebrate animals. Sequencing was carried out (Christensen et al., 2018) using the tissues of the female IW2-2015 obtained from the company engaged in industrial aquaculture of chars – Icy Waters Ltd. The company exploits two of its own broodstocks – NL and TR, originating from the chars from the Nauyuk Lake and the Tree River (Nunavut, Canada). Since the complete mitochondrial genome of the female IW2-2015 was absent in the published assembly ASM291031v2, we determined its type and complete sequence from the sequence read archives taken from Genbank. It was found that the female's mitogenome belongs to the BERING haplogroup, which is characteristic of Northern Dolly Varden *S. malma malma*. Analysis of other unlinked diagnostic loci encoded by nuclear DNA (ITS1, RAG1, SFO-12, SFO-18, SMM-21) also revealed distinctive characters of Northern Dolly Varden in female IW2-2015. It was concluded that the genomic assembly ASM291031v2 was obtained not from an individual of Arctic char *S. alpinus*, but from an individual of a related species – Northern Dolly Varden *S. malma malma*. The identical to the IW2-2015 female characteristics of diagnostic loci were found in other individuals from the broodstock TR. Apparently, the broodstock TR is entirely a strain derived from Northern Dolly Varden. The choice of a specific specimen for sequencing and assembling the genome of a species is no less important than the choice of a species, since both of these choices determine the directions for subsequent data use. Assembly ASM291031v2, of course, is important for the study of the genomic architecture of the aquaculture strain TR. However, since it was obtained from a specimen originated from the marginal population (Tree River) isolated from the main range of the species and with some traces of introgressive hybridization, this assembly can hardly be considered as a description of a typical genome of Northern Dolly Varden.

Supplementary Data: https://github.com/svshedko/CHARgenome

*УДК 575.11+575.1]:597.552.51*

СБОРКА ASM291031v2 (GENBANK: GCA_002910315.2) ИДЕНТИФИЦИРОВАНА КАК СБОРКА ГЕНОМА СЕВЕРНОЙ МАЛЬМЫ (*Salvelinus malma malma*), А НЕ ГЕНОМА АРКТИЧЕСКОГО ГОЛЬЦА (*S. alpinus*)

С. В. Шедько

*Федеральный научный центр Биоразнообразия наземной биоты Восточной Азии Дальневосточного отделения Российской академии наук, Владивосток, 690022*

e-mail: shedko@biosoil.ru

АННОТАЦИЯ

У лососевых рыб к настоящему моменту секвенировано двенадцать полных геномов, представляющих одиннадцать видов, принадлежащим шести родам. В роде *Salvelinus* предполагалось секвенировать геном арктического гольца – одного из самых изменчивых видов позвоночных животных. Секвенирование проведено (Christensen et al., 2018) с использованием тканей самки IW2-2015, полученной от компании Icy Waters Ltd, занимающейся промышленной аквакультурой гольцов. Icy Waters Ltd в производстве эксплуатирует два собственных маточных стада NL и TR, ведущих свое начало от гольцов из оз. Нуаяк и р. Три (провинция Нунавут, Канада) соответственно. Поскольку в опубликованной сборке ASM291031v2 полный митохондриальный геном самки IW2-2015 отсутствовал, то его тип и полная последовательность были установлены нами из архивов прочтений, взятых из Genbank. В результате оказалось, что митогеном самки IW2-2015 относится к гаплогруппе BERING, характерной для северной мальмы *S. malma malma*. Анализ других несцепленных диагностических локусов, кодируемых ядерной ДНК (ITS1, RAG1, SFO-12, SFO-18, SMM-21), также выявил у самки IW2-2015 отличительные черты северной мальмы. Сделан вывод, что геномная сборка ASM291031v2 получена не от особи арктического гольца *S. alpinus*, а от особи близкого вида – северной мальмы *S. malma malma*. Такие же, как у самки IW2-2015, состояния характеристик диагностических локусов были найдены и у других особей из маточного стада TR. По-видимому, маточное стадо TR целиком является линией, производной от северной мальмы. Выбор конкретной особи для секвенирования и сборки генома какого-либо вида не менее важен, чем выбор вида, поскольку оба этих выбора определяют направления последующего использования данных. Сборка ASM291031v2, безусловно, имеет важное значение для исследования геномной архитектуры аквакультурной линии TR. Однако, поскольку она была получена от особи, происходящей из краевой популяции, изолированной от основного ареала вида и с некоторыми следами произошедшей интрогрессивной гибридизации, эта сборка вряд ли может рассматриваться как описание типичного генома северной мальмы.

*Ключевые слова*: ядерный геном, митохондриальный геном, сборка, Salmonidae.

Гольцы рода *Salvelinus* – одна из эволюционных линий лососевых рыб сем. Salmonidae, распространенная в Северном полушарии, преимущественно в его высоких широтах. Многие из видов гольцов издавна привлекали внимание исследователей своим исключительным морфологическим разнообразием, проявляющимся не только при сравнении географически изолированных популяций, но и при рассмотрении их внутри одной и той же водной системы. Гольцы – одни из доминантов в бедных видами северных ихтиоценозах. Ведя в основном проходной образ жизни, гольцы, тем не менее, не способны переносить низкую зимнюю температуру вод с морской соленостью. Поэтому большую часть года они проводят в пресных водах, оказываясь при этом доступными местному населению как источник ценной деликатесной пищи.

Арктический голец *S. alpinus* – циркумполярно распространенный вид, в котором отличительные черты гольцов проявляются наиболее ярко. Арктический голец по праву может считаться одним из самых изменчивых видов позвоночных животных (Klemetsen et al., 2003; Klemetsen, 2010, 2013). Его ареал ближе к северному полюсу, чем ареал любого из видов пресноводных или проходных рыб. Арктический голец имеет важное значение как объект спортивного рыболовства, промысла и аквакультуры (Johnston, 2006). Поэтому не удивительно, что среди всех видов *Salvelinus* первым в очереди на секвенирование и аннотацию полного генома оказался именно арктический голец.

Для секвенирования генома арктического гольца, его сборки и аннотации (Christensen et al., 2018) была использована небольшая (20 см) самка (её идентификатор – IW2-2015), полученная от компании Icy Waters Ltd, занимающейся промышленной аквакультурой гольцов. Компания является одним из крупнейших производителей продукции (мясо-филе, икры) из искусственно выращенных гольцов для реализации на рынках Северной Америки. В своем производстве Icy Waters Ltd использует два маточных стада гольцов – NL и TR, ведущих свое начало от 15–25 особей, взятых соответственно из природных популяций гольцов оз. Нуаяк и р. Три, расположенных на арктическом побережье Канады в зал. Коронации, регион Китикмеот, провинция Нунавут (Goel, 2004; Johnston, 2006; McGowan et al., 2009).

При ознакомлении со сборкой генома арктического гольца ASM291031v2 и сопровождающей её публикацией (Christensen et al., 2018) обратило на себя внимание следующее противоречие. В статье на второй странице указано, что гольцы из оз. Нуаяк и р. Три имеют митохондриальную ДНК (мтДНК), относящуюся к гаплогруппе ARCTIC. Однако, в Genbank в качестве сборки неядерного генома арктического гольца (RefSeq: GCF_002923155.1) приведен митогеном из гаплогруппы ACADIA (GenBank: AF154851.1=NC_000861.1), полученный ранее в другом исследовании (Doiron et al., 2002) и для других гольцов с северо-востока Квебека (Канада).

Проведенное нами в попытке прояснить эту ситуацию картирование прочтений геномной ДНК (Genbank: SRX3776048–SRX3776052), а также прочтений РНК-секвенирования (Genbank: SRX2635048, SRX2635052) на диагностические гаплотипы мтДНК гольцов показало, что митогеном самки IW2-2015 однозначно принадлежит к мтДНК-гаплогруппе BERING, характерной, как известно, не для арктического гольца, а для северной мальмы – *S. malma malma* (Brunner et al., 2001). Другие привлеченные диагностические маркеры также свидетельствовали о том, что самка IW2-2015 скорее всего является представителем северной мальмы, а не арктического гольца. Результаты этого анализа изложены ниже.

## МАТЕРИАЛ И МЕТОДИКА

В качестве материала была использована сборка генома самки IW2-2015 ASM291031v2 (GenBank: GCA_002910315.2), а также архивы прочтений её ДНК (GenBank: SRX3776048–SRX3776052) и РНК из тканей печени и предпочки (GenBank: SRX2635048, SRX2635052).

Кроме того, анализировались архивы прочтений геномной ДНК, полученные для двух особей гольцов из маточного стада TR компании Icy Waters Ltd – SRX5282523, SRX5282524 и нескольких комбинаций гибридов племенных линий TR и NL: TR♂ × NL♀ – SRX5282528, SRX5282529; (TR♂ × NL♀)♂ × NL♀ – SRX5282525; TR♂ × ((TR♂ × NL♀)♂ × TR♀)♀ – SRX5282526.

Идентификация принадлежности особей, послуживших источником геномных данных, осуществлялась с использованием шести маркеров: мтДНК и пяти локусов ядерной ДНК – ITS1, RAG1, SFO-12, SFO-18, SMM-21.

У альпиноидных (арктический голец и близкие или производные от него формы) и мальмоидных (северная мальма *S. malma malma* и близкие или производные от неё формы) гольцов по структуре контролирующего региона мтДНК ранее было выделено шесть гаплогрупп – ATLANTIC, SIBERIA, ACADIA, ARCTIC, BERING (Bruner et al., 2001) и OKHOTSKIA (Shedko et al., 2007). У арктического гольца Северо-Западных Территорий и провинции Нунавут распространена гаплогруппа ARCTIC при абсолютном доминировании в выборках гаплотипа ARC19 (Moore et al., 2015). У северной мальмы на Аляске и Северо-Западных Территориях, как и в азиатской части ареала, господствует гаплогруппа

BERING. Доля гаплотипа BER12 из этой гаплогруппы в выборках часто доходит до 50 и более процентов, как, например, в выборках мальмы с Командорских о-вов или Камчатки [(Soshnina et al., 2016): гаплотип под номером KT962126]. Гаплотипы ARC19 (Genbank: EU310899) и BER12 (Genbank: JX261984) различаются по девяти из 499 нуклеотидных позиций маркерного участка контролирующего региона мтДНК (восемь из них – диагностические для гаплогрупп в целом). Эти два гаплотипа выступили в качестве референсных при экспресс-диагностике типа мтДНК-гаплогрупп в геномных данных. Для сборки митогеномов целиком прочтения картировали на соответствующие варианты полных митогеномов: ARC19 – MF621741 (Genbank) и BER12 – KJ746618 (Genbank).

Первый внутренний транскрибируемый спейсер (ITS1) является составной частью тандемных повторов рибосомальной ДНК (рДНК), располагающейся у альпиноидных и мальмоидных гольцов на одной или нескольких парах хромосом (Phillips et al., 1999). Альпиноидным и мальмоидным (исключая южную американскую мальму *S. malma lordi*) гольцам присущи разные варианты ITS1, отличающиеся недвусмысленными замещениями в позициях 92 (C/T), 285 (T/A), 448 (C/A), 449 (T/A), 450 (G/C) и 453 (A/C) (Phillips et al., 1999). Координаты приведены относительно последовательности AF059893 (Genbank), выбранной как маркерная для альпиноидного варианта. Для идентификации мальмоидного варианта ITS1 в этом качестве выступила последовательность AF059900 (Genbank).

В геномах лососевых рыб ген RAG1 (ген активирующий рекомбинацию 1) присутствует в виде одной копии. Этот ген у лососевых имеет сравнительно низкую скорость эволюции, но несет в себе ясный филогенетический сигнал (Shedko et al., 2012a). Альпиноидным и мальмоидным гольцам свойственны разные варианты гена RAG1 (а именно – фрагмента второго экзона, длиной 1524 пн), отличающиеся недвусмысленными замещениями в позициях 285 (G/A) и 1113 (G/A) (PopSet:306977732). Координаты даны относительно последовательности гена RAG1 северной мальмы *S. malma malma* GQ871481 (Genbank), выбранной в качестве референсной.

Исследование состава аллелей в микросателлитном локусе SFO-12 у разных видов гольцов из их различных популяций показало (Angers, Bernatchez, 1997; Wilson, Bernatchez, 1998), что арктическому гольцу (подчеркнем, что в первой из указанных работ имелись образцы непосредственно из оз. Нуаяк) свойственен аллель SFO-12(225), а северной мальме *S. malma malma* и её сателлитным формам (белому гольцу р. Камчатки, например) аллель SFO-12(223). Различие этих аллелей определяется разным числом TG-повторов по обе стороны от GG-димера:

(TG)$_4$GG(TG)$_8$ и (TG)$_5$GG(TG)$_6$ соответственно. Из-за этого идентификация этих двух аллелей получается вполне однозначной, несмотря на минимальную разницу в их длине. Последовательности этих аллелей вместе с фланкирующими их участками были восстановлены исходя из рис. 1–2 работы (Angers, Bernatchez, 1997). Их длина в результате оказалась равной 221 (для аллеля, обозначенного в цитированной работе как 223) и 223 (для аллеля под кодом 225). Такие же их длины получены при анализе геномных данных.

Аллели микросателлитного локуса SFO-18 (155–157, с одной стороны, и 161–164, с другой) дифференцируют мальму и арктического гольца с Аляски и востока арктического побережья Канады (Angers, Bernatchez, 1996; Crane et al., 2014; Ditlecadet et al., 2006; Hart et al., 2015; Taylor, May-McMally, 2015). Аллели различаются числом повторов CA-мотива у базовых вариантов – 7 у мальмы против 10 у арктического гольца (Genbank: MN530967, MN530968). Популяции арктического гольца провинции Нунавут необычны тем, что в них примерно с равной частотой встречаются оба этих аллеля (Harris et al., 2016).

Микросателлитный локус SMM-21 считается еще одним диагностическим маркером для арктического гольца и северной мальмы. Для арктического гольца и близких к нему форм характерны аллели длиной 105–111 пн (Crane et al., 2014; Гордеева и др., 2010; Сенчукова, 2014; Hart et al., 2015; Taylor, May-McMally, 2015), а для северной мальмы – 115–158 пн (Crane et al., 2005, 2014; Салменкова и др., 2009; Сенчукова, 2014; Harris et al., 2015; Hart et al., 2015; Taylor, May-McMally, 2015). Различия в длинах вызваны вариацией числа повторов TC-мотива (Genbank: AY327128, MN530969–MN530972).

Установление вариантов аллелей шести диагностических локусов в геномных данных производилось следующим образом.

Последовательности гаплотипов контролирующего региона ARC19 и BER12, двух вариантов ITS1 (AF059893 и AF059900), аллелей 223 и 221 локуса SFO-12, а также гена RAG1 были объединены в один файл, который выступал в качестве составного референса. Для локусов SFO-18 и SMM-21 референсы были индивидуальными. Прочтения из того или иного архива выравнивались на референс с помощью программы blastn, доступной на сайте https://blast.ncbi.nlm.nih.gov. Выравнивание выполнялось в режиме "megablast" (поиск последовательностей с высоким уровнем сходства) (Zhang et al., 2000) с ограничением максимального числа целевых последовательностей в 1000 шт. Данный метод прост в использовании и имеет подходящую для нашего случая чувствительность и необходимый уровень избирательности. Полученный на выходе программы blastn SAM-файл затем анализировался с помощью программы Tablet v.1.19.05.28 (Milne et al., 2013), и по характеру покрытия и типу нуклеотидов в диагностических позициях устанавливался вариант аллелей маркерных локусов. Для уточнения megablast-выравнивания по локусу SMM-21 выровненные прочтения извлекались и картировались на референс вновь, но уже с помощью программы Bowtie 2 v2.3.2 (Langmead, Salzberg, 2012).

В тех случаях, когда требовалось максимально полно использовать архивные прочтения (для проверки megablast-выравнивания, оценки покрытия или последующей сборки последовательности целевого локуса), первичные данные загружались из European Nucleotide Archive (www.ebi.ac.uk/ena) в виде FASTQ-файлов, содержащих разбитые по направлениям прочтения. Затем с помощью утилиты bbduk из пакета BBMap производилась фильтрация данных на предмет остаточного присутствия адаптеров, использованных при секвенировании. Далее осуществлялось выравнивание с помощью утилиты bbmap из пакета BBMap v.38.00 (Bushnell, 2014) с ключами "local" и "vslow" или программы Bowtie 2 в режиме "local" или "end-to-end". При необходимости, выровненные прочтения в формате FASTQ или FASTA служили вводными данными для сборки целевых последовательностей с помощью программы SPAdes v3.11.1 (Nurk et al., 2013) или idba_ud v1.0.9 (Peng et al., 2012).

Для ресурсоёмких операций и расчетов использовали возможности многопроцессорного вычислительного комплекса IRUS17 Центра коллективного пользования "Дальневосточный вычислительный ресурс" ДВО РАН (г. Владивосток).

РЕЗУЛЬТАТЫ И ОБСУЖДЕНИЕ

Результаты анализа геномных данных по всему использованному набору маркерных локусов представлены в таблице.

Как указано во введении, аутентичная для самки IW2-2015 полная последовательность мтДНК в сборке ASM291031v2 отсутствует. Blastn-анализ не выявил также в этой сборке корректно собранных участков рибосомальной ДНК. Наибольшее сходство с референсной последовательностью ITS1 демонстрировал участок NC_036862.1:28683274–28682712 (Genbank). Но уровень этого сходства был относительно невелик (95–96%). Если к этому референсу добавляли фланкирующие его участки 18S и 5.8S РНК, то уровень его сходства с выровненным участком (Genbank: NC_036862.1:28683274–28682712) падал до 91%. В тоже время, последовательности ITS1, полученные при анализе архивов прочтений геномной ДНК или РНК (все три найденных варианта ITS1 вместе с прилегающими участками генов 18S и 5.8S РНК были депонированы в Genbank: MN530964–MN530966), имели 100%-ное сходство с уже известными для гольцов (таблица). Отсюда можно понять, что в сборке ASM291031v2 качественно собранные участки рДНК, по существу, отсутствуют.

Варианты последовательностей четырех других маркерных локусов (RAG1, SFO-12, SFO-18 и SMM-21), установленные в результате blastn-анализа сборки ASM291031v2, в каждом случае идентифицированы как аллели, свойственные северной мальме.

Анализ архивов прочтений геномной ДНК самки IW2-2015, взятых из Genbank, выявил следующую картину. Самка IW2-2015 имеет мтДНК из гаплогруппы BERING (гаплотип – BER12). Последовательности RAG1, SFO-18 и SMM-21, в полном согласии с результатами анализа сборки ASM291031v2, относятся к варианту северной мальмы.

К варианту северной мальмы относится и подавляющее большинство прочтений для локуса ITS1. В качестве минорного компонента среди последних выступил альпиноидный вариант А2. Стоит подчеркнуть, что, судя по результатам анализа прочтений, полученных в ходе РНК-секвенирования (Genbank: SRX2635048, SRX2635052), экспрессируется из них лишь один вариант – мальмоидный.

По локусу SFO-12 у самки IW2-2015 найдены оба варианта аллелей (и альпиноидный, и мальмоидный). В двух экспериментах (Genbank: SRX3776048–SRX3776049) преобладает вариант мальмоидных гольцов, в других (Genbank: SRX3776050–SRX3776052) – их соотношение примерно равное.

Анализ архивов прочтений, полученных для двух особей из маточного стада TR (Genbank: SRX5282523–SRX5282524), привел к результатам, полностью идентичным результатам анализа самки IW2-2015. То есть, судя по ним, самка IW2-2015 безусловно относится к линии TR.

Из архивов прочтений SRX3776048 и SRX5282523 были собраны два варианта полных митохондриальных геномов (Genbank: MN530959, MN530961), характеризующих линию TR. Эти митогеномы отличались друг от друга по двум нуклеотидным позициям, а от митогеномов мальмоидных гольцов Камчатки (Genbank: KJ746618, KU674351, KU674352, KT266871, KT266870) – по 30–39 позициям.

У гибридов TR♂ × NL♀ (Genbank: SRX5282528–SRX5282529) найден гаплотип мтДНК из гаплогруппы ARCTIC, а по ядерным локусам – альтернативные варианты аллелей примерно в равной пропорции. Характеристики одного из вариантов возвратных скрещиваний (Genbank: SRX5282526) оказались близки к характеристикам линии TR (к слову, полный митогеном этой особи был полностью идентичен митогеному, восстановленному из архива SRX5282523), а другого (Genbank: SRX5282525) – к характеристикам актического гольца. Отметим, что в последнем случае, как и у гибридов TR♂ × NL♀, был найден особый вариант ITS1 – альпиноидный вариант А3. Из архивов прочтений SRX5282525 и SRX5282528 были восстановлены два варианта полных митогеномов из гаплогруппы ARCTIC (Genbank: MN530963, MN530962), различавшихся двумя нуклеотидными позициями. Варианты этих митогеномов, а также аллелей ядерных локусов, полученные для архива SRX5282525, по всей видимости, могут рассматриваться как характеристики, близкие к характеристикам (если не идентичные им) особей из маточного стада NL.

По результатам проведенного анализа, видно, что характеристики племенных линий TR и NL повторяют характеристики северной мальмы и арктического гольца соответственно. Другими словами, эти линии, по всей видимости, происходят от особей, принадлежавшим двум разным видам гольцов. В связи с этим, можно утверждать, что опубликованная сборка генома самки IW2-2015 (ASM291031v2, Genbank: GCA_002910315.2) является сборкой генома северной мальмы *S. malma malma*.

Река Три в зал. Коронации, откуда получены особи-основатели линии TR, одно из известных [особенно среди любителей спортивной гольцовой рыбалки из-за возможности отметиться крупным трофеем весом до 14 кг (Moshenko et al., 1984)] местообитаний мальмы в регионе Китикмеот провинции Нунавут (Alfonso et al., 2018). Для мальмы этот район является пределом её распространения на восток по арктическому побережью Северной Америки. Заметим, что здесь проходит восточная граница ареалов многих других видов тихоокеанских морских или анадромных рыб (Alfonso et al., 2018): тихоокеанской сельди *Clupea pallasii*, кеты *Oncorhynchus keta*, нерки *O. nerka*, чавычи *O. tshawytscha*, азиатской зубастой корюшки *Osmerus dentex*, тихоокеанской наваги *Eleginus gracilis*, звездчатой камбалы *Platichthys stellatus*, дальневосточной зубатки *Anarhichas orientalis*. Таким образом, следует иметь ввиду то, что сборка ASM291031v2 характеризует геном особи, происходящей из краевой популяции, изолированной от основного ареала вида, заканчивающегося бассейном р. Маккензи.

Краевые популяции часто малочисленны и из-за этого могут быть подвержены гибридизации с популяциями близких видов. Весьма вероятно, что присутствие альпиноидного варианта А2 ITS1, а также аллеля SFO-12(223) в геномах особей из линии TR является следами произошедшей гибридизации северной мальмы и арктического гольца в регионе Китикмеот. Широкое присутствие аллеля SFO-18(156) у арктического гольца из этого региона (Harris et al., 2016) также может быть следствием такой гибридизации. Поэтому геном самки IW2-2015 не может рассматриваться как описание типичного генома северной мальмы. С другой стороны, этот геном не является показательным и для арктического гольца, поскольку предковые линии мальмоидных и альпиноидных гольцов разошлись довольно давно – 3.03 (1.18–5.00) млн лет назад (Shedko et al., 2012b). В тоже время, полученные в работе (Christensen et al., 2018) данные, безусловно, имеют важное значение для исследований геномной архитектуры аквакультурных TR- и NL-линий этих видов гольцов.

Подчеркнем, что полноту и качество существующей сборки ASM291031v2 пока, к сожалению, нельзя признать высокими. Об этом свидетельствует, во-первых, то, что в работе (Christensen et al., 2018) не был собран митохондриальный геном и, во-вторых, то, что в сборке ASM291031v2 отсутствуют корректные последовательности рибосомальной ДНК.

Наличие адекватного референсного генома – базовое условие всех проводимых на геномном уровне исследований в той или иной группе организмов (Elmer, 2016). Здесь, безусловно, важен не только выбор вида в качестве типичного представителя этой группы, но и выбор конкретного индивидуума, представляющего сам вид. В идеале эта особь должна происходить из популяции, типичной для этого вида, и располагающейся в центральной части его ареала. Хочется надеяться, что в будущих попытках секвенирования геномов гольцов (мальмы, арктического гольца или других) этому обстоятельству будет уделено особое внимание.

исследования в интересах комплексного развития Дальневосточного отделения РАН" (грант №18-4-040).

ПРИЛОЖЕНИЯ

Все значимые нуклеотидные последовательности, восстановленные из архивов прочтений, депонированы в Genbank под номерами доступа MN530959–MN530972.

Основные исходные данные, использованные для построения таблицы, а также другие материалы можно найти в репозитории, доступном по ссылке https://github.com/svshedko/CHARgenome.

СПИСОК ЛИТЕРАТУРЫ

**Таблица**. Варианты аллелей маркерных локусов, характерные для арктического гольца и северной мальмы и установленные для самки IW2-2015 (в фигурных скобках – соотношение альтернативных вариантов прочтений в случае полиморфизма), двух особей из маточного стада TR, а также четырёх гибридных особей, полученных в нескольких вариантах скрещивания племенных линий TR и NL.
**Table**. Alleles of diagnostic loci characteristic for Arctic char and Northern Dolly Varden, revealed in female IW2-2015 (in curly brackets is the ratio of alternative reads in the case of polymorphism), two individuals from broodstock TR, as well as four variants of hybrids.

| Таксон/образец/данные (номера доступа Genbank) | мтДНК | ITS1[1] | RAG1 | SFO-12 | SFO-18 | SMM-21 |
|---|---|---|---|---|---|---|
| *S. alpinus* | ARCTIC | alpinus | alpinus | 223 | 162–164 | 105–111 (alpinus) |
| *S. malma malma* | BERING | malma | malma | 221 | 156 | 115–158 (malma) |
| **IW2-2015** | | | | | | |
| GCA_002910315.2 | – | – | malma[2] | 221[3] | 156[4] | malma[5] |
| SRX3776048 | BER12 | A2/malma {~1:55} | malma | 221/223 {~13:1} | 156 | malma |
| SRX3776049 | BER12 | A2/malma {~1:34} | malma | 221/223 {~6:1} | 156 | malma |
| SRX3776050 | BER12 | A2/malma {~1:55} | malma | 221/223 {~1:1} | 156 | malma |
| SRX3776051 | BER12 | A2/malma {~1:235} | malma | 221/223 {~1:2} | 156 | malma |
| SRX3776052 | BER12 | malma | malma | 221/223 {~1:1} | 156 | malma |
| SRX2635048, SRX2635052 | BER12 | malma | malma | – | – | – |
| **TR** | | | | | | |
| SRX5282523 | BER12 | A2/malma {~1:30} | malma | 221/223 {~1:1} | 156 | malma |
| SRX5282524 | BER12 | A2/malma {~1:30} | malma | 221/223 {~1.5:1} | 156 | malma |
| **Гибриды** | | | | | | |
| SRX5282528 TR♂ × NL♀ | ARC19 | A3/malma {~1:2} | alpinus[6]/malma {~1:3} | 221/223 {~2:1} | 156/162 {~2:1} | alpinus/malma {~2:1} |
| SRX5282529 TR♂ × NL♀ | ARC19 | A3/malma {~1:2} | alpinus[6]/malma {~1:1} | 221/223 {~2:1} | 156/162 {~1.4:1} | alpinus/malma {~4:1} |
| SRX5282525 (TR♂ × NL♀)♂ × NL♀ | ARC19 | A3/malma {~5:1} | alpinus[6] | 223 | 156/162 {~2:1} | alpinus |
| SRX5282526 TR♂ × ((TR♂ × NL♀)♂ × TR♀)♀ | BER12 | A2/malma {~1:33} | malma | 221/223 {~1.5:1} | 156/162 {~2:1} | alpinus/malma {~3:1} |

*Примечания.* [1] A2 – альпиноидный вариант (Genbank: AF059899); A3 – альпиноидный вариант (Genbank: AF059897), найден у арктического гольца из оз. Нуаяк (Philllips et al., 1999), отличается уникальной для гольцов нуклеотидной заменой C>T в позиции 298. [2] NC_036842.1:69870375–69871898 (Genbank), LG4q.1:29. [3] NC_036863.1:41094706–41094926 (Genbank), LG23. [4] NC_036875.1:27971989–27972144 (Genbank), LG36. [5] NC_036838.1:46248006–46248134 (Genbank), LG1. [6] Редкий для альпиноидных гольцов вариант RAG1 (Genbank: GQ871484), до сих пор был найден только у боганидской палии из оз. Эльгыгытгын на Чукотке (Shedko et al., 2012a).

*Notes.* [1] A2 – alpinoid variant (Genbank: AF059899); A3 – alpinoid variant (Genbank: AF059897), found in Arctic char from Nauyuk Lake (Philllips et al., 1999), is distinguished by a unique for alpinoid chars C>T nucleotide substitution in position 298. [2] NC_036842.1:69870375–69871898 (Genbank), LG4q.1:29. [3] NC_036863.1:41094706–41094926 (Genbank), LG23. [4] NC_036875.1:27971989–27972144 (Genbank), LG36. [5] NC_036838.1:46248006–46248134 (Genbank), LG1. [6] A rare for alpinoid chars variant of RAG1 (Genbank: GQ871484), so far it has been found only in Boganida char from Lake El'gygytgyn in Chukotka (Shedko et al., 2012a).